\documentclass{PoS}

\title{The physics of the radio emission in the quiet side of the AGN
  population with the SKA}

\ShortTitle{Radio emission in radio-quiet AGN}

\author{\speaker{M. Orienti}$^{1}$\thanks{orienti@ira.inaf.it},
F. D'Ammando$^{1,2}$, M. Giroletti$^{1}$, G. Giovannini$^{1,2}$,
F. Panessa$^{3}$\\
$^1$INAF -- IRA, Bologna, Italy \\
$^{2}$University of Bologna, Bologna, Italy\\
$^{3}$INAF-IAPS, Roma, Italy\\ 
}

\abstract{Super massive black holes (SMBH) 
are thought to be ubiquitously hosted in
massive galaxies. They may be either quiescent, like the case of
Sgr\,A* in our Galaxy, or active, and they are at the basis of the
phenomena known as Active Galactic
Nuclei (AGN). In this case they often manifest their presence by
releasing a huge amount of energy which usually overwhelms the star-related 
contribution of the entire host galaxy. Although they have been targets of many
multiwavelength campaigns, the main physical processes at work in
AGN are still under
debate. In particular the origin of the radio emission and the
mechanisms involved are among the open questions in astrophysics. 
The radio-loud AGN population and their radio
emission is linked to the presence of bipolar outflows of relativistic
jets. However, the large majority of the AGN population do not form
powerful highly-relativistic jets on kpc scales, like those
  observed in radio galaxies and radio quasars.
This does not mean that they are
radio-silent objects. On the contrary, these systems are characterized
by radio luminosity up to 10$^{23}$ W/Hz at 1.4 GHz, challenging our
knowledge on the physical processes at the basis of the radio 
emission in radio-quiet objects. 
The main mechanisms proposed so far are synchrotron
radiation from mildly relativistic mini-jets, thermal
cyclo-synchrotron emission by low-efficiency accretion flow (like ADAF
or ADIOS), or thermal free-free emission from the X-ray heated corona
or wind. The difficulty in understanding the main mechanism involved is
related to the weakness of these objects, which precludes the study of
non-local radio-quiet AGN.
Multifrequency, high-sensitivity polarimetric radio
observations are, thus, crucial to constrain the nature of the power
engine, and they may help in distinguishing between the contribution
from star formation and AGN activity.
The advent of the Square Kilometer Array (SKA), with its sub-arcsecond
resolution and unprecedented sensitivity 
will
allow us to investigate these processes in radio-quiet AGN, even at high
redshift for the first time. Both the broad-band radio spectrum and
the polarization information will help us in
disentangling between non-thermal and thermal origin of the radio emission.  
The jump in sensitivity of a few order of
magnitudes at the (sub-)$\mu$Jy level will enable us to detect
radio emission from a large number of radio-quiet AGN at high
redshift, providing a fundamental step in our understanding of their
cosmological evolution.
}

\FullConference{
Advancing Astrophysics with the Square Kilometre Array\\
June 8-13, 2014\\
Giardini Naxos, Italy}

\begin{document}

\section{Introduction}
 Galaxies represent the majority of the observable matter in the
 Universe. Gravitationally bound gas, dust and stars are their basic
 constituents. In their centre a super massive black
 hole (SMBH) is usually found, and if it is in an ``active''
 phase it is responsible for a significant
contribution to the SED that cannot be attributed to other origins
such as stars, gas and dust. 
In this case, the central region of the galaxy is termed
 an Active Galactic Nucleus (AGN) and is one of the most energetic
 phenomena in the Universe. Its emission is observed
  in a large range of the electromagnetic spectrum, from infrared (IR)
  to X-rays, by processes directly and indirectly related to the
  release of gravitational 
  energy from matter falling onto the SMBH. \\
If the AGN is able to form
  a bipolar outflow of relativistic plasma, then its radio emission becomes
  comparable to or even stronger than the emission observed in the other
  energy bands. The presence, or lack, of relativistic jets is at the
  basis of the radio-loud and radio-quiet dichotomy. Only 10 per cent
  of the AGN population is radio-loud, while in the large majority the
  radio emission is a negligible part of the bolometric
  luminosity. The latter are 
  termed radio-quiet AGN and their radio luminosity at 1.4 GHz
  does not exceed 10$^{23}$ W/Hz (e.g. Condon et al. 1992). \\
Although the radio emission is a very marginal part of the energy
released by radio-quiet AGN, it represents a unique way to
investigate the high energy particle accelerators.
Understanding the origin of the radio emission from radio-quiet AGN is
not trivial. Usually 
radio-quiet AGN are hosted in late-type galaxies where star formation
is responsible for the majority of the radio emission. For this
reason, disentangling the AGN-related emission from the stellar
contribution is a hard task to perform. The knowledge of the physical
processes occurring in the nuclear region of radio-quiet AGN is a
critical point for understanding differences and similarities with the
radio-loud AGN population, as well as for investigating a possible
link between the AGN and the star formation. \\
The advent of the
Square Kilometer Array (SKA) will provide a jump forward in the
observational capabilities reaching unexplored sensitivity levels, and
thus allowing the study of faint and non-local objects,
improving the statistics and providing a fundamental step forward in
our understanding of the physical processes at work and their
implications for the AGN feedback.\\
In this Chapter we
briefly introduce the key issues about the radio emission in the
radio-quiet AGN population and we discuss how
the advent of the SKA will help in shedding a light on this hot topic.
The Chapter is organized as follows: in Section 2 we present the state
of the art of the radio emission in radio-quiet AGN; in Section 3 we
indicate how SKA may address some open questions during the SKA1
phase, and the jump that is expected when the array will be fully
operative. Concluding remarks are in Section 4.\\

\section{Radio emission in radio-quiet AGN: state-of-the-art}

\subsection{The stellar contribution}

A significant fraction of the radio emission in radio-quiet AGN comes
from processes related to the stellar evolution, like synchrotron
emitting cosmic rays accelerated by supernovae, and thermal free-free
radiation from the ionized gas in star forming regions. 
The fact that the tight radio/far-infrared (FIR)
correlation found for star-forming galaxies (SFG) holds in radio-quiet
AGN supports the idea that the bulk of
the radio emission in radio-quiet AGN is related to stellar processes,
while the agreement is poor in radio-loud AGN due to the presence of
relativistic jets (e.g. Padovani et al. 2011).\\
When observed with adequate angular resolution and sensitivity 
the radio emission is often spread
across the host galaxy. A clear example is represented by the
Seyfert galaxy NGC\,1097. On arcsecond scale the radio emission 
traces the profile of the host galaxy and its spiral
arms (Hummel et al. 1987, Condon 1987), while 
on smaller scale the radio emission is
organized in a spectacular circumnuclear starburst ring 
where several star forming
regions are clearly identified. \\
Circumnuclear starburst emission enshrouding the central AGN is observed
in many Seyfert galaxies, like Circinus (Elmouttie et al. 1998), NGC\,7469
and NGC\,7586 (e.g. Orienti \& Prieto 2010). The co-spatial distribution of the
AGN and starburst activity has suggested a connection between
these two phenomena. A close link between star formation and 
radio emission in radio-quiet AGN is further supported by their evolution
which is indistinguishable from that observed in SFG
and their luminosity function (LF), which appears to be an
extension of the 
SFG LF (Padovani et al. 2011).\\

\subsection{The particle accelerators in the nuclei of radio-quiet AGN}

The origin of the radio emission from the nucleus of radio-quiet AGN
is a matter of debate. The fact that the FIR flux better correlates
with the low-resolution kpc-scale radio flux density, rather than the
high-resolution pc-scale emission disfavours a dominant stellar origin
for the nuclear emission (Thean et al. 2000).\\
The main mechanisms proposed include 1) thermal
emission/absorption from hot gas (Gallimore et al. 2004); 2) 
low-efficiency accretion/radiation flow (Narayan \& Yi 1994, Blandford
\& Begelman 1999); 3)
non-thermal synchrotron radiation from a mildly relativistic jet 
(Orienti \& Prieto 2010), or 4) a combination of processes
(Falcke \& Markoff 2000, Ghisellini et al. 2004). \\
The study of a sample of local Seyfert nuclei pointed out an empirical
correlation between radio and X-ray luminosities, suggesting that the
accretion flow and the radio emission are strongly coupled
(Panessa et al. 2007). Interestingly, this correlation seems the same as
the one found for radio-loud radio galaxies, whose radio emission is
related to relativistic jets. This may indicate a possible common
mechanism between the two populations. However, this claim cannot be
unambiguously proved due to the large difference in luminosity
between Seyferts and radio galaxies. \\
%
%
Support for the presence of synchrotron
emission from the Seyfert nuclei comes from the detection of radio
structures similar to those observed in powerful radio source
(i.e. jets, lobes and hot spots), like
the case of Circinus (Elmouttie et al. 1998) and
NGC\,1068 (Ulvestad et al. 1987). 
In these two sources, evidence for AGN able to
accelerate particles to high energy was claimed after the detection by
the Large Area Telescope on board {\it Fermi} of
$\gamma$-ray emission which seems to exceed the cosmic-ray
contribution from the host galaxy (Hayashida et al. 2013, Lenain et
al. 2010).\\
In Seyfert galaxies the radio emission is confined within the host
  galaxy, whereas radio galaxies have radio structures on scales 
up to hundred of kiloparsecs or even megaparsecs. 
Not all Seyfert nuclei behave in the same way. In particular,
it has been found that flat-spectrum Seyfert nuclei usually do not
show extended morphology, and the radio emission comes from an
unresolved (sub-)parsec-scale region (Anderson \& Ulvestad 2005).
On the contrary, in the steep-spectrum
Seyfert nuclei the radio emission is not centrally concentrated, but
rather is diffuse over a larger region
(e.g. NGC\,4151, Ulvestad et al. 1998).
In some sources kpc-scale bubbles are observed (e.g. Mrk\,6,
  Mingo et al. 2011). These bubbles may
drive shocks in the interstellar medium of the host galaxy and may play
a role in regulating the star formation in the hosts (e.g. Mingo et
al. 2012).
These differences may arise from different physical mechanisms: 
steep-spectrum nuclei may be able to produce extended, but slow jet
structures, whereas in flat-spectrum nuclei the energy released by the
AGN is mainly localized in the innermost region, without developing
jets.\\

\section{The role of SKA}

Due to its weakness, the radio emission of radio-quiet AGN has been
investigated mainly in nearby objects. If the resolution is not
adequate, the nuclear component may be washed out by the
stellar-related emission, and the radio properties may be
contaminated by the contribution from different components. \\
VLBI observations have turned out to be an effective hunter of
AGN by detecting 
compact, variable components with brightness temperatures above
10$^{6}$ K, and high core dominance (i.e. the ratio between the
milliarcsecond and arcsecond flux density). 
Panessa \& Giroletti (2013) studied a complete sample of local radio-quiet
AGN by collecting heterogeneous VLBI observations, confirming the
importance of deep high-resolution observations on more statistically
complete samples and at different redshift.
%
%
This issue becomes even more complicated when the sub-mJy
population is taken into account due to observational limitations.\\
A correct determination of the stellar- and AGN-related emission is
important for a comprehensive characterization of the radio emission
from the AGN, its evolution and the possible interplay with the host
galaxy. These key issues can be addressed by:

\begin{itemize}

\item determining the core dominance by comparing low-resolution and
  high-resolution radio observations, which provides information on
  the fraction of the radio emission concentrated in the central
  region, 

\item the study of the broad-band radio spectrum
and the polarization properties
  of the nuclear region, which are important tools for testing
    whether the radio emission is synchrotron radiation.

\end{itemize}

The expected SKA performances will provide a jump forward in our
understanding of the physics of these extraordinary celestial
bodies.\\

\subsection{Unveiling the nuclear emission with SKA1-MID}

A primary requirement for the study of the AGN contribution to the radio
emission is an adequate angular resolution. With a baseline of about
100 km, the SKA1-MID array has the potential to address this
issue. The angular resolution that will be achieved ranges between $\sim$0.4
arcsec at 21 cm (band 2) and $\sim$0.07 arcsec at 3.6 cm (band
5). This resolution has already proved to be adequate for separating the
AGN emission from the possible contribution of nuclear star forming
regions. In fact, if we consider the bulk of the radio-quiet AGN at $z
\sim 1.5-2$,
a resolution of 0.07 arcsec corresponds to $\sim$0.6 kpc, while the
host galaxy should have a total angular size of $\sim$2
arcsec. The separation of the nuclear emission becomes of course
easier as we consider closer objects.
In addition, a minimum baseline of $\sim$400 m will allow the
detection of the diffuse emission in nearby objects up to $\sim$1.5
arcmin (band 2), 
enabling a proper spatial characterization of both star-forming processes and
possible extended jet structures.\\
The availability of multifrequency polarimetric observations will be
crucial for unveiling both the spectral index and polarization
distribution. 
Assuming a continuum sensitivity (770-MHz band) of $\sim$0.57$\mu$Jy
h$^{-1/2}$ (natural weight, Table 1 in Dewdney et al. 2013), it would be
possible to study large sample of radio-quiet AGN with a 10$\sigma$ 
sensitivity of $\sim$10 $\mu$Jy (uniform weight) in a reasonable
time. The high resolution coupled with the 
deep sensitivity will allow one to reliably separate the AGN emission
from the stellar contribution. This would allow the study of radio
emission from AGN with
luminosity $L\sim10^{18}$ W/Hz, $L\sim10^{20}$ W/Hz, $L\sim5\times
10^{22}$ W/Hz, and 
$L\sim10^{23}$W/Hz at redshift 0.01, 0.1, 1, and 2, respectively. The
two-tiered survey at Band 5 will be a starting point for this study.\\
The availability of a (quasi-)continuous radio spectrum will be a
crucial tool for addressing this issue. In fact, thermal and
synchrotron self-absorbed spectra are expected to show different
properties, like the slope and the location of the peak
frequency. \\
Observations covering a continuous frequency range from $\sim$ 1 GHz
to $\sim$ 10 GHz will be crucial for discriminating between the
thermal and non-thermal radio emission. This would require that band 2
(950-1760 MHz), 4
(2800-5180 MHz) and 5 (4600-13800 MHz) should be installed in SKA1-MID. 
A flat spectrum up to high frequency will be a clear evidence for
thermal emission, while a turnover around a few GHz strongly
supports the non-thermal synchrotron radiation. Furthermore, if the
peak of the spectrum is due to synchrotron self-absorption, the
frequency is strongly related to the physical properties of the
emitting region: $\nu_p \propto H^{1/5} B^{2/5}$, where $H$ is the
magnetic field and $B$ is the brightness. If the magnetic field
computed from the observational parameter is unrealistically high,
then it would prove that the peak in the spectrum is not due to
synchrotron self-absorption, like in the case of NGC\,4457 where the
estimated magnetic field was $\sim$10$^{9}$ G (Bontempi et al. 2012).\\
A spectral resolution of 100 MHz would be adequate for a proper
characterization of the radio spectrum. Assuming the performances
expected for the SKA1-MID ($\sim$63 and 82 $\mu$Jy h$^{-1/2}$ for a 100 kHz
spectral resolution in band 2 and 5, respectively), it would be
possible to easily achieve a 5$\sigma$ 
sensitivity of $\sim$20 $\mu$Jy with a 100-MHz spectral resolution in
about one hour (uniform weight). \\
The broad-band radio spectrum may be combined with the polarization
information for constraining the nature of the radio
emission arising from different regions. For example, a jet structure
is expected to show polarized steep-spectrum synchrotron emission,
while the diffuse synchrotron emission from a star-forming 
region should be 
highly depolarized due to the tangled magnetic field caused by the
supernovae explosions. However, low-resolution observations may cause
beam depolarization in case of non-resolved jet structures.\\
A more complex issue will be understanding the physical processes at
the basis of the nuclear, 
flat-spectrum emission from the core region. In principle one may
expect that unpolarized emission may originate in
thermal bremsstrahlung radiation from the gas of the hot corona, while
some (very low) level of polarization may be observed in presence of
synchrotron self-absorption. However, we must keep in mind that strong
depolarization from the gas 
enshrouding the central region is likely to play a major
role. Therefore, the lack of polarized emission cannot reliably
discriminate between the two processes.\\

{\bf The SKA1-MID early-science capabilities}\\

Important results on local AGN 
would be already achievable during the early-science
operations. Assuming a maximum baseline of $\sim$ 50 km, and the
availability of frequencies up to 5 GHz (band 4), the resolution would
be 0.2 arcsec. This would preclude us from resolving the central
kiloparsec region beyond redshift $z \sim$0.4. However, objects with
$z < 0.1$ would be studied in excellent details and 
we should already be able to build the
(quasi-)continuous spectrum of the different regions with a 5$\sigma$
threshold of $\leq$60 $\mu$Jy in a reasonable time. 
This would allow the study of objects
with a luminosity of $L\sim10^{21}$ W/Hz at redshift $z=0.1$, down
to luminosity $L\sim10^{19}$ W/Hz at $z=0.01$.\\

\subsection{Towards the entire array: SKA}

The advent of the full SKA array will provide a jump in the
observational capabilities. The deployment of the receivers up to 24
GHz will provide a step forward in the study of the nuclear emission of
either thermal or non-thermal origin. For a maximum baseline of
2000 km (i.e. 20 times longer than the SKA1-MID) the resolution at
high frequency will be a few milliarcsecond, allowing the
characterization of parsec-scale regions even at high redshift. This
will be accompanied by an improvement of about one order of magnitude
of the sensitivity, allowing the unprecedented detection of (sub-)$\mu$Jy objects 
with largely affordable exposure time. \\
In addition to the study of the total intensity emission from the
nuclear region, observations allowing the detection
of circular polarization may be a fundamental tool for identifying
cyclo-synchrotron emission from low-efficiency accretion/radiation
flow.\\

\section{Concluding remarks}

The radio-quiet AGN population is expected to represent a large
fraction of the faint radio sky that will be picked up by the Square
Kilometer Array. Due to its weakness, the radio emission from these
objects is not fully understood. Furthermore, the co-existence of both
the AGN and star-formation activity makes the segregation of these two
components a hard task. Constraining the radio properties of both
contributions will be fundamental for determining the mechanisms at
work, how they evolve,
constraining the unbiased luminosity functions, and exploring a
possible interplay between them.\\
The observational capabilities of SKA will be crucial for addressing
these key issues:\\

\begin{itemize}

\item disentangling the contribution of the star-formation activity
  from the nuclear emission in nearby as well as in high redshift
  objects. This will be achieved by the availability of high angular
  resolution and deep sensitivity,

\item understanding the nature of the radio emission from the central
  AGN. The quasi-continuous radio spectra covering a large range of
  frequencies will help in discriminating thermal bremsstrahlung
  radiation from hot gas, from non-thermal synchrotron radiation,

\item determining the presence of extended jet-like structure related
  to the AGN activity by the analysis of the polarized emission.

\end{itemize}

\noindent The advent of the Square Kilometer Array will provide a substantial
advance in our understanding of the radio-quiet AGN population, and
the role that they play in the evolution of the host galaxy.\\

\noindent {\bf References}\\
 

\noindent Anderson, J.M., Ulvestad, J.S. 2005, ApJ, 627, 674


\noindent Blandford, R.D.; Begelman, M.C. 1999, MNRAS, 303, L1


\noindent Bontempi, P., Giroletti, M., Panessa, F., Orienti, M., Doi, A. 2012,
MNRAS, 426, 588


\noindent Condon, J.J. 1987, ApJS, 65, 485


\noindent Condon, J.J. 1992, ARA\&A, 30, 575


\noindent Dewdney, P., Turner, W., Millenaar, R., McCool, R., Lazio,
J., Cornwell, T., 2013, ``SKA1 System Baseline Design'', Document number
SKA-TEL-SKO-DD-001 Revision1


\noindent Elmouttie, M. Haynes, R.F., Jones, K.L., Sadler, E.M., Ehle, M. 1998, MNRAS,
297, 1202


\noindent Fabian, A.C., Rees, M.J. 1995, MNRAS, 277, L55


\noindent Falcke, H., Markoff, S. 2000, A\&A, 362, 113


\noindent Gallimore, J.F., Baum, S.A. O'Dea, C.P. 2004, ApJ, 613, 794


\noindent Ghisellini, G., Haardt, F., Matt, G. 2004, A\&A, 413, 535


\noindent Hayashida, M., Stawarz, L., Cheung, C.C., et al. 2013, 779, 131 


\noindent Ho, L.C., Ulvestad, J.S. 2001, ApJS, 133, 77


\noindent Hummel, E., van der Hulst, J.M., Keel, W.C. 1987, A\&A, 172, 32



\noindent Lenain, J.-P., Ricci, C., T\"urler, M., Dorner, D., Walter, R. 2010,
A\&A, 524, 72

\noindent Mingo, B., Hardcastle, M.J., Croston, J.H., et al. 2011, ApJ, 731, 21

\noindent Mingo, B., Hardcastle, M.J., Croston, J.H., et al. 2012, ApJ, 758, 95

\noindent Nagar, N.M., Wilson, A.S., Falcke, H. 2001, ApJ, 559, 87


\noindent Narayan, R., Yi, I. 1994, ApJ, 428, 13


\noindent Orienti, M., Prieto, M.A. 2010, MNRAS, 401, 2599


\noindent Padovani, P., Miller, N., Kellermann, K.I., Mainieri, V., Rosati, P., Tozzi, P.
2011, ApJ, 740, 20


\noindent Panessa, F., Barcons, X., Bassani, L., et al. 2007, A\&A, 467, 519

\noindent Panessa, F., Giroletti, M. 2013, MNRAS, 432, 1138


\noindent Thean, A., Pedlar, A., Kukula, M.J., Baum, S.A., O'Dea, C.P. 2000,
MNRAS, 314, 573


\noindent Thean, A., Pedlar, A., Kukula, M.J., Baum, S.A., O'Dea, C.P. 2001,
MNRAS, 325, 737


\noindent Ulvestad, J.S., Neff, S.G., Wilson, A.S. 1987, AJ, 93, 220


\noindent Ulvestad, J.S., Roy, A.L., Colbert, E.J.M., Wilson, A.S. 1998, ApJ,
496, 196


\end{document}